%%%%%%%%%%%%%%%%%%%%%%%%%%%%%%%

%%%%%%%%%%%%%%%%%%%%%%%%%%%%%%%%%%%%%%%%%%
\input phyzzx
\hfuzz 15pt

\def\dplus{=\hskip-5pt \raise 0.7pt\hbox{${}_\vert$} ^{\phantom 7}}
\def\dplusup{=\hskip-5.1pt \raise 5.4pt\hbox{${}_\vert$} ^{\phantom 7}}
%%%%%%%%%%%%%%%%%%%%%%%%%%%%%%%%%%%%%%%%%%%%%%%%%%%%%%%%%%%%%%%%%%
\def\dplus{=\hskip-4.8pt \raise 0.7pt\hbox{${}_\vert$} ^{\phantom 7}}

\def\pmb#1{\setbox0=\hbox{#1} \kern-.025em\copy0\kern-\wd0
\kern0.05em\copy0\kern-\wd0 \kern-.025em\raise.0433em\box0}

\def\hT{\hat T}

\font\mybb=msbm10 at 10pt

\def\bb#1{\hbox{\mybb#1}}

\def\bC {\bb{C}}

\def\bR{\bb {R}}

\def\a{\alpha}
\def\b{\beta}
%%%%%%%%%%%%%%%%%%%%%%%%%%%%%%%%%%%%%%%%%%%%%%%%%%%%%%%%%%%%%%%%%%%%
\REF\cmh{S.J. Gates, C.M. Hull and M. Ro\v cek, Nucl. Phys. {\bf B248}
(1984) 157.}
\REF\hpa {P.S. Howe and G. Papadopoulos, Nucl .Phys. {\bf B289} (1987) 264;
Class. Quantum Grav. {\bf 5} (1988) 1647; Nucl .Phys. {\bf B381} (1992) 360.}
\REF\hkt {P.S. Howe and G. Papadopoulos, Phys. Lett. {\bf B379} (1996) 80.}
\REF\ish{S. Ishihara, J. Diff. Geom. {\bf 9} (1974) 483.}
\REF\sal{S. Salamon, Invent Math. {\bf 67} (1982), 143; Ann. Sci.
 ENS Supp. {\bf 19} (1986).}
\REF\nishino{H. Nishino, Phys. Lett. {\bf B355} (1995) 117.}
\REF\pvn{B. de Wit and P. van Nieuwenhuizen, Nucl. Phys. {\bf B312} (1989) 58.}
\REF\swann{ A. Swann, C.R. Acad. Sci. Paris, t. {\bf 308} (1989) 225.}
\REF\hpb{P.S. Howe and G. Papadopoulos, Commun. Math. Phys. {\bf 151}
(1993) 467.}
\REF\hw{C.M. Hull and E. Witten, Phys. Lett. {\bf 160B} (1985) 398.}
\REF\sen{A. Sen, Nucl. Phys. {\bf B278} (1986) 289. }
\REF\hpc{ P.S. Howe and G. Papadopoulos, Class. Quantum Grav.
{\bf 4} (1987) 1749.}
\REF\eb{ E. Bergshoeff and E. Sezgin, Mod. Phys. Lett. A Vol 1 (1986) 191.}
\REF\chs{C.G. Callan, J.A. Harvey and A. Strominger, Nucl. Phys. 
{\bf B359} (1991) 611.}
\REF\lebrun{C. LeBrun, Journ. of Diff. Geom. {\bf 34} (1991) 223.} 
\REF\gio{A. Galperin, E. Ivanov and O. Ogievetsky,
 Ann. Phys. {\bf 230} (1994) 201.}
\REF\roc{N.J. Hitchin, A. Karlhede, U. Lindstr\"om, and M. Ro\v cek,
 Commun. Math. Phys. {\bf 108}
(1987) 535.}
\REF\dks{F. Delduc, S. Kalitsin and E. Sokatchev, Class. Quantum. Grav. {\bf 7}
(1990) 1567.}
\REF\wolf{J. Wolf, J. Math. Mech. {\bf 14} (1965) 1033.}
\REF\vp {Ph. Spindel, A. Sevrin, W. Troost and A. Van Proeyen, 
Nucl. Phys. {\bf B308} (1988) 662;
{\bf B311} (1988) 465.}
\REF\gps{G.W. Gibbons, G. Papadopoulos and K.S. Stelle, 
{\sl HKT and OKT geometries on soliton
black hole moduli spaces}, hep-th/9706207.}
\REF\ao{A. Opfermann, {\sl T-duality and HKT manifolds}, hep-th/9709048.}

%%%%%%%%%%%%%%%%%%%%%%%%%%%%%%%%%%%%TITLE PAGE%%%%%%%%%%%%%%%%%%%%%%
\Pubnum{ \vbox{ \hbox{R/97/50}\hbox{KCL-TH-97-56} } }
\pubtype{}
\date{October, 1997}
\titlepage
\title{Twistor spaces for QKT manifolds}
\author{P.S. Howe}
\address{Department of Mathematics \break King's College London\break London
WC2R 2LS}
\andauthor{A. Opfermann and G. Papadopoulos}
\address{\negthinspace DAMTP \break Silver Street \break Cambridge CB3 9EW}
%\address{}
\abstract {We find that  the target space of two-dimensional (4,0)
supersymmetric sigma models with torsion coupled to (4,0) supergravity 
is a QKT manifold, that is, a quaternionic K\"ahler manifold with torsion.   
We  give four 
examples of geodesically complete QKT manifolds one of which is a
 generalisation of the LeBrun
geometry.  We then construct the twistor space associated with
 a QKT manifold and show that under
certain conditions it is a K\"ahler manifold with a complex 
contact structure. We also show that,
for every 4k-dimensional QKT manifold, there is an 
associated 4(k+1)-dimensional hyper-K\"ahler
one. }

\endpage
\pagenumber=2
%%%%%%%%%%%%%%%%%%%%%%%%%%%%%%%%%%%%MACROS%%%%%%%%%%%%%%%%%%%%%%%%%
\font\mybb=msbm10 at 12pt
\def\bb#1{\hbox{\mybb#1}}

\def\CN {\bb{C}}

\def\C{\mkern1mu\raise2.2pt\hbox{$\scriptscriptstyle|$}\mkern-7mu{\rm C}}

\def\a{\alpha}
\def\b{\beta}

\sequentialequations
%%%%%%%%%%%%%%%%%%%%%%%%%%%%%%%INTRODUCTION%%%%%%%%%%%%%%%%%%%%%%%%%%%%%%%%

\chapter{Introduction}

The geometry of the target space of two-dimensional
 sigma models with extended supersymmetry
is described by the properties of a metric connection
 with torsion [\cmh, \hpa]. Rigid
(4,0) supersymmetry requires that the target space of 
 two-dimensional sigma models without
Wess-Zumino term (torsion) is a hyper-K\"ahler (HK)
 manifold. In the presence of torsion the
geometry of the target space becomes hyper-K\"ahler with 
torsion (HKT) [\hkt]. Manifolds with either
HK or HKT structure  admit three complex structures which
 obey the algebra of imaginary unit
quaternions and  the sigma model metric is hermitian
with respect to all complex structures.  In
addition, in  HK geometry the complex structures are
 covariantly constant with
respect to the Levi-Civita connection, while in  HKT
geometry the complex
structures are covariantly constant with respect to
 a metric connection with torsion. 
Local (4,0) supersymmetry requires that the target
 space of two-dimensional sigma
models with torsion be either (i) HKT or (ii) a 
generalisation of the standard
quaternionc K\"ahler geometry (QK) 
(see [\ish, \sal]) for which the
associated metric connection has torsion 
[\nishino]; we shall call this geometry
quaternionic K\"ahler with torsion (QKT). 
This is unlike the case of (4,4) locally
supersymmetric sigma models  where it has 
been shown that the geometry of the
target space is either of HKT type or it 
is standard  quaternionic K\"ahler geometry
[\pvn].  Thus QKT  geometry is not compatible 
with (4,4) local
supersymmetry.  Nevertheless, the conditions on the
 geometry of the target space of
two-dimensional sigma models required by (4,0)-local
 supersymmetry can be derived from
an appropriate truncation of the conditions found 
for the  (4,4)-locally
supersymmetric ones.   It is well known in QK
 geometry that the holonomy of the
Levi-Civita connection is a subgroup of
$Sp(k)\cdot Sp(1)$. Similarly,  QKT geometry is
characterised by the fact that the holonomy of
 a metric connection with torsion has
holonomy  $Sp(k)\cdot Sp(1)$.  The torsion is
the exterior derivative of the Wess-Zumino term
 of the sigma model action and
is therefore a closed three-form on the sigma
 model manifold, at least in the classical theory.

In this paper, we list the conditions on the
 target manifold of a sigma model
required by (4,0)-local supersymmetry and thus 
derive the  restrictions
on a Riemannian manifold that must be satisfied 
in order for it to admit a QKT geometry.  We shall
then explore some of the properties of QKT geometry. In
particular, we shall show that for every four-dimensional
 quaternionic K\"ahler manifold
there is an associated class of QKT manifolds. These
 manifolds are parameterised by
harmonic functions (possibly with delta function
 singularities on the QK manifold). This
gives a large class of QKT manifolds since every 
orientable 4-manifold is QK due to the fact that
$SO(4)=Sp(1)\cdot Sp(1)$.  Using this method, we present
 four examples of complete
four-dimensional QKT manifolds. Allowing $dH\not=0$, we
 show that any $4k$-dimensional QKT
manifold admits a twistor construction. We construct
 the twistor space of a QKT manifold and
show that
 it is a
K\"ahler manifold with a complex contact structure
 provided that $k>1$ and $dH$ is (2,2)-form with
respect to all three complex structures.  In addition, 
we associate to every $4k$-dimensional
QKT manifold a $4(k+1)$-dimensional HK one which is a fibre
bundle over the QKT manifold with fibre $\bC^2-{0}$. 
In the limit that the torsion $H$ vanishes the
results  of  Salamon  [\sal] and  Swann [\swann] 
for QK manifolds are recovered.

This paper is organised as follows:  in section two
 we state the algebraic and
differential conditions required by QKT geometry on a
 Riemannian manifold; in section three we present  
examples of QKT manifolds; in section four we
give the twistor construction for QKT manifolds and
 show that the twistor space is K\"ahler with a
complex contact structure; in section five we show 
that for any QKT manifold there is an associated
HK one, and  in section six we make some concluding remarks.

%%%%%%%%%%%%%%%%%%%%%%%%%%%%%%%%%%%%%%%%%%%%%%%%%%%%%%%%%

\chapter{Local (4,0) supersymmetry}

The multiplets required for the construction of a
 two-dimensional (4,0) locally supersymmetric
theory coupled to sigma model matter are as follows: 
(i) The supergravity multiplet $(e, C, \psi)$
comprises of the graviton
$e$, a $SO(3)$ gauge field $\{C^r; r=1,2,3\}$ and 
four Majorana-Weyl gravitini $\{\psi, \psi^r$;
$r=1,2,3\}$; (ii) sigma model scalar multiplets 
$(\phi, \chi)$, each comprised of four real scalars
$\phi$ and four Majorana-Weyl fermions $\chi$.  
The spinors of the scalar multiplet have
opposite chirality to those of the supergravity one. Let
$M$ be the sigma model manifold of dimension
$4k$ with metric
$g$,  Wess-Zumino 3-form $H$,  a ${\cal L}SO(3)$-valued
 one-form $B$ and three almost complex
structures $\{I_r; r=1,2,3\}$.
 The Lagrangian  \foot{The letters from the beginning
 of the Greek alphabet
$\a,\b, \gamma, \delta=0,1$ are worldvolume induces
  and the letters from the middle of the Greek
alphabet are target space indices
$\lambda, \mu, \nu, \kappa=1,\dots, 4k$. We have also 
suppress spinor indices.} that describes the
(4,0)-supergravity multiplet coupled to $k$ scalar
 multiplets system is
$$
\eqalign{
2 e^{-1}{\cal L}&=g_{\mu\nu} \partial_\a \phi^\mu \,
\partial^\a \phi^\nu-\epsilon^{\a\b}
b_{\mu\nu}\partial_\a \phi^\mu \,\partial_\b \phi^\nu
\cr &
-ig_{\mu\nu} \bar\chi^\mu \gamma^\a{\cal D}_\a  \chi^\nu +
ig_{\mu\nu} \bar\chi^\mu \gamma^\a \gamma^\b\partial_\b\phi^\kappa
\big(\delta_\kappa{}^\nu \psi_{\a}- (I_r)_\kappa{}^\nu \psi^r_{\a}\big) 
\cr &
-{1\over3} \bar\chi^\lambda\gamma^\a\chi^\nu 
\bar\chi^\mu \big(3 H_{\kappa[\lambda \nu}
(I_r){}_{\mu]}{}^\kappa
\psi^r_{\a }+ H_{\mu\lambda \nu} \psi_{\a }\big)
\cr &
-{1\over8} g_{\mu\kappa}\bar\chi^\mu\gamma^\a
\chi^\nu \big( (I_r){}_\nu{}^\lambda 
\bar\psi^r_{\b } +
\delta_\nu{}^\lambda \bar\psi_{\b }\big)\gamma_\a \gamma^\delta
\gamma^\b\big((I_r){}_\lambda{}^\kappa \psi^r_{\delta } -
\delta_\lambda{}^\kappa \psi_{\delta } \big)\ , }
\eqn\appone
$$
where
$$
{\cal D}_\a\chi^\mu=\nabla^{(+)}_\a\chi^\mu+ B_\a{}^r 
(I_r){}_\nu{}^\mu \chi^\nu- {1\over2}
\omega_\a \chi^\mu- C_\a{}^r (I_r){}_\nu{}^\mu \chi^\nu\ ,
\eqn\apptwo
$$
$B^r_\a$ is the pull back of $B^r_\mu$ with respect 
to $\phi$, $\omega_\a$ is the spin
connection of the worldvolume and the 
covariant derivatives $\nabla^{(\pm)}$ are
 associated with the connections
$$
\Gamma^{(\pm)}_{\nu\kappa}{}^\mu=\hat\Gamma_{\nu\kappa}{}^\mu
\pm{1\over2}H_{\nu\kappa}{}^\mu\, ;
\eqn\inatwo
$$
$\hat\Gamma$ is the Levi-Civita connection of the metric $g$. 
To simplify the notation we set
$\Gamma=\Gamma^{(+)}$ ($\nabla=\nabla^{(+)}$).

The conditions on the geometry of $M$ required by (4,0)
 local supersymmetry can be 
found by appropriately truncating  the conditions 
required by  (4,4) local supersymmetry
[\pvn]. The former are the following:
$$
\eqalign{
I_r I_s&=-\delta_{rs}+\epsilon_{rst} I_t
\cr
(I_r){}_\mu{}^\kappa (I_r){}_\nu{}^\lambda
 g_{\kappa \lambda} &=g_{\mu\nu}\ ; \qquad r=1,2,3
\cr
D_\mu (I_r){}_\kappa{}^\nu&=0\cr
{\cal N}_{\hat D}(I_r){}^\mu{}_{\nu\kappa}&=0\ ,}
\eqn\intone
$$
where
$$
D_\mu (I_r){}_\kappa{}^\nu=\nabla_\mu
 (I_r){}_\kappa{}^\nu+B_\mu{}_r{}^s
(I_s){}_\kappa{}^\nu\ ,
\eqn\inttwo
$$
$B_r{}^s= -2 B^t \epsilon_{tr}{}^s$. In addition 
$$
{\cal N}_{\hat D}\big(I_r\big){}^\mu{}_{\nu\kappa}=(I_r){}_\nu{}^\lambda {\hat D}_{[\lambda}
(I_{r}){}_{\kappa]}{}^\mu- (\nu\leftrightarrow \kappa)
\eqn\inafour
$$
is a Nijenhuis-like tensor  associated with the covariant derivative
$$
{\hat D}_\mu (I_r){}_\nu{}^\kappa=\hat \nabla_\mu (I_r){}_\nu{}^\kappa+ B_{\mu\, r}{}^s
(I_s){}_\nu{}^\kappa\ ,
\eqn\new
$$
where $\hat \nabla$ is the Levi-Civita covariant derivative.  
We remark that this Nijenhuis tensor is 
independent from the Levi Civita part of ${\hat D}$.

The first three conditions in \intone\ imply that (i) the 
almost complex structures, $I_r$,
obey the algebra of imaginary unit quaternions, (ii) the
 metric $g$ is hermitian with
respect to all almost complex structures and (iii) the
 holonomy of the connection $D$ is a
subgroup of $Sp(k)\cdot Sp(1)$, respectively.  The 
covariantised Nijenhuis condition,  ${\cal
N}_D(I_r)=0$, together with the third condition 
in \intone\ imply that the torsion is (1,2)-and
(2,1)-form with respect to all almost complex structures. 
 We remark that in the commutator of two
supersymmetry transformations, apart from
${\cal N}_{\hat D}(I_r)$,  the mixed covariantised Nijenhuis 
`tensors', ${\cal N}_{\hat D}(I_r,
I_s)$, appear as well (see for example [\hpb]).  However 
they do not give independent conditions on
the almost complex structures since
 they vanish provided that ${\cal N}_{\hat D}(I_r)=0$.

In analogy with the HKT case [\hkt],  we say that 
the manifold $M$ with tensors $g,I, B$ and 
$H$ that satisfy \intone\ has a weak QKT structure
 if  no further conditions are
imposed on $H$. However, if in addition we take $H$
 to be a closed 3-form
($dH=0$), we say that $M$ has a strong QKT structure, 
in which case we can write
$$
H=3 db
\eqn\inasix
$$
for some locally defined two-form $b$ on $M$.  Finally, 
if $H$ vanishes,
the manifold $M$ becomes quaternionic K\"ahler. The 
target space, $M$, of a
(classical) (4,0) locally supersymmetric sigma model
 with torsion is a manifold with a strong QKT
structure. The couplings of the classical action of
 the theory are the metric
$g$, the ${\cal L}SO(3)$ valued one-form $B$ 
and the two-form $b$. 
However, in the quantum theory  and in particular
 in the context of the anomaly cancellation
mechanism [\hw,\sen, \hpc], the (classical) torsion $H$ of
(2,0)-supersymmetric sigma models receives corrections\foot{Apart
from the sigma model anomalies, these models have also  two-dimensional
gravitational anomalies.}.  The new torsion is not a closed three form.
 Therefore, although
classically the target space of (4,0)-supersymmetric 
sigma models has a strong QKT structure,
quantum mechanically this may change to a
 weak QKT structure, albeit of a
particular type.

It is well known that  all $4k$-dimensional QK manifolds are 
Einstein, i.e.
$$
R_{\mu\nu}=\Lambda g_{\mu\nu}\ ,
\eqn\inaseven
$$
and that
$$
G^r_{\mu\nu}=\big(dB+ B\wedge B\big)^r{}_{\mu\nu}
=-{ \Lambda\over k+2} (I^r){}_{\mu\nu}\ ,
\eqn\inaeight
$$
where $\Lambda$ is a constant and $G$ is the
 curvature of the $B$ connection. There
is no direct analogue of these statements in
 the context of QKT geometry.  However, one
can show that {\sl if} the curvature $G$ of 
the $B$ connection satisfies \inaeight\
then the torsion $H$ vanishes. To show this,
 we first differentiate \inaeight\ with
respect to the $\nabla_\kappa$ connection
 and then antisymmetrise in all three
$i,j,k$ indices.  Then, using the fact that 
$DI_r=0$ we find that the right hand side of
the equation can be expressed in terms of $B$ and $I_r$. 
Using \inaeight\ once more we find that the left
hand side of the equation is expressed in terms of the torsion and a term similar
to that of the right hand side. Finally, one gets
$$
 (I_r){}_{[\kappa}{}^\lambda H_{ \mu\nu]\lambda} =0\ .
\eqn\inanine
$$ 
Using this together with the fact that H is
 a (2,1) and (1,2) tensor on $M$, we
conclude that $H$ vanishes.  Thus equation
 \inaeight\ excludes torsion. A
consequence of this is that the (4,0) locally
 supersymmetric models constructed in
[\eb] have zero torsion. 

%%%%%%%%%%%%%%%%%%%%%%%%%%%%%%%%%%%%%%%%%%%%%%%%%%%%%%%%%%%%

\chapter{Examples}

To construct examples of QKT geometry, we generalize
 the ansatz used in [\chs] to
find HKT geometries from HK ones.  As we 
have already mentioned in the
introduction, any oriented four-dimensional manifold  is QK.  Let
$M$ be such a manifold with metric $h$, connection $B$ and compatible
almost complex structures $I_r$.  The volume form, 
$\Omega$, of $M$ can be written in
terms of the almost complex structures as 
$$
\Omega=\sum^3_{r=1} \omega_r\wedge \omega_r\ ,
\eqn\exone
$$
where 
$$
\omega_r(X,Y)= h(X,I_rY)
\eqn\extwo
$$
are the K\"ahler-like forms of the almost complex
 structures $I_r$. We remark that
$\Omega$ is covariantly constant with respect to 
the Levi-Civita connection. We also
mention for later use that\foot{Although 
the 4-form $\Omega$ can be defined for
 QK manifolds of any dimension, this identity 
holds only for four-dimensional manifolds.}
$$
\sum^3_{r=1} (\omega_r)_{\mu\nu}\, (\omega_r)_{
\kappa\lambda}=\Omega_{\mu\nu \kappa\lambda}+
h_{\mu\kappa} h_{\nu\lambda}-h_{\nu\kappa} h_{\mu\lambda}\ .
\eqn\exthreea
$$

To construct four-dimensional QKT manifolds, we
use the ansatz
$$
g=F\, h, \qquad H={1\over2}\star dF\, ,
\eqn\exthree
$$
where $\star$ is the Hodge dual with respect to $\Omega$.
 The manifold $M$ with
metric $g$, torsion $H$, almost complex structures
 $I_r$ and connection $B$  is a weak
QKT manifold.  To show the covariant constancy 
condition of the almost complex
structures in \intone, we use the equation \exthreea\ and the ansatz
 \exthree. The remaining
conditions in
\intone\ are straightforwardly  satisfied.    
For $M$ to have a strong QKT
structure, $H$ must be closed which in turn
 implies that $F$ must be a harmonic
function on $M$ with respect to the $h$ metric, i.e.
$$
d\star dF=0\ .
\eqn\exthreeb
$$
We shall allow $F$ to have delta function singularities on $M$.  
There  always exist non-trivial solutions of 
\exthreeb\ on any four-dimensional manifold.
So we conclude that there is a family of 
QKT manifolds associated  to every
four-dimensional QK manifold labeled by the
 harmonic functions of the latter\foot{Note
that the QKT manifold with metric $g$ is also
 QK with respect to the same metric, as
four-dimensional manifold, but with  a different
 set of almost complex structures.}.

Due to the singularities of $F$, the associated 
QKT metric may be geodesically
incomplete. This in fact is the case for 
some choices of harmonic function for the
compact four-dimensional Wolf spaces $S^4$ 
and  $\CN P^2$.  However there are examples
of complete QKT geometries. Here we shall
 present four non-singular QKT manifolds
starting from the QK manifolds, 
$\bR\times dS_{(3)}$, $dS_{(4)}$, the Tolman
wormhole and a LeBrun like metric, 
respectively, where $dS_{(n)}$ is n-dimensional de Sitter
space.    

The metric $h$ on $\bR\times dS_{(3)}$ is
$$
ds^2= du^2+ dv^2+\cosh^2v\, d\Omega^2_{(2)}\ , 
\eqn\rds
$$
where $-\infty <u, v<\infty$ and $d\Omega^2_{(2)}$
 is the $SO(3)$ invariant metric on $S^2$. 
Supposing that the harmonic function, $F$, depends only on $v$, we get
$$
F=\lambda_1 \tanh(v)+\lambda_2\ ,
\eqn\rdsa
$$
where $\lambda_1$ and  $\lambda_2$ are real numbers.
It is straightforward to compute the metric
 and the torsion of the QKT manifold to find
that
$$
\eqalign{
ds^2_F&=(\lambda_1 \tanh(v)+\lambda_2)\, 
\big[du^2+ dv^2+\cosh^2v\, d\Omega^2_{(2)} \big]
\cr
H&= \lambda_1\sin\theta\, du\wedge d\theta\wedge d\phi\ ,}
\eqn\rdsb
$$
where $\theta, \phi$ are the angular coordinates on $S^2$. 
This QKT metric is geodesically complete 
if we choose $\lambda_2>|\lambda_1|$.

The metric $h$ on $ dS_{(4)}$ is
$$
ds^2= dv^2+\cosh^2v\, d\Omega^2_{(3)} \ , 
\eqn\ds
$$
where $-\infty < v<\infty$ and 
$d\Omega^2_{(3)}$ is the $SO(4)$ invariant metric on $S^3$. 
Supposing that the harmonic function, 
$F$, depends only on $v$, we get 
$$
F=\lambda_1 \big[{\sinh(v)\over \cosh^2(v)}+
{\rm arctan}(\sinh(v))\big]+\lambda_2\ ,
\eqn\dsa
$$
where $\lambda_1$ and $\lambda_2$ are real numbers.
It is straightforward to compute the metric
 and the torsion of the QKT manifold to find
that
$$
\eqalign{
ds^2_F&=\big(\lambda_1 \big[{\sinh(v)\over \cosh^2(v)}+
{\rm arctan}(\sinh(v))\big]+\lambda_2\big)\
\big[dv^2+\cosh^2v d\Omega^2_{(3)}\big]
\cr
H&= {\lambda_1\over2}\sin^2\theta\, \sin\phi\,  d\theta\wedge d\phi\wedge d\psi\ ,}
\eqn\dsb
$$
where $\theta, \phi, \psi$ are the angular coordinates on $S^3$.  
This QKT metric is geodesically complete, 
if we choose $\lambda_2>{\pi\over2} \lambda_1>0$.

The metric of the Tolman wormhole  is
$$
ds^2= dv^2+(a^2+v^2) d\Omega^2_{(3)}\ , 
\eqn\wh
$$
where $-\infty < v<\infty$, $a$ is a real non-zero
 constant, and $d\Omega^2_{(3)}$ is the
$SO(4)$ invariant metric on $S^3$. This metric is
the analytic continuation of the FRW
model of a universe filled with a perfect fluid
 with pressure  equal to $1/3$ of its density. 
Using the Einstein equations, we find the Tolman wormhole metric has
zero scalar curvature. In addition, it is conformally flat
$$
ds^2= \big(1+{a^2\over 4 r^2}\big)^2 (dr^2+ r^2\, d\Omega^2_{(3)})\ ,
\eqn\wha
$$
 as can be easily seen using the coordinate transformation
$$
v=r-{a^2\over 4 r}\ .
\eqn\whb
$$
Supposing that the
harmonic function,
$F$, depends only on
$v$, we get
$$
F=\lambda_1\, {v\over a^2 \sqrt {a^2+v^2}}+\lambda_2\ ,
\eqn\rdsa
$$
where $\lambda_1$ and $\lambda_2$ are real numbers.
It is straightforward to compute the metric and the
 torsion of the QKT manifold to find
that
$$
\eqalign{
ds^2_F&=\big(\lambda_1\, {v\over a^2 \sqrt {a^2+v^2}}
+\lambda_2\big)\, \big[ dv^2+(a^2+v^2)
d\Omega^2_3\big] 
\cr
H&= {\lambda_1\over2}\sin^2\theta\, \sin\phi\,  d\theta\wedge d\phi\wedge d\psi\ ,}
\eqn\rdsb
$$
where $\theta, \phi, \psi$ are the angular coordinates on $S^3$. 
This QKT metric is geodesically complete, 
if we choose $\lambda_2>(1/a^2) |\lambda_1|$.

All the examples of QKT geometries presented so far 
are conformally flat and therefore their Weyl
tensor vanishes. For reasons that will become apparent
 in the twistor construction of four-dimensional
QKT manifolds, we give an example of a  QKT geometry with non-vanishing but
self-dual Weyl tensor. To do this we begin with the four-dimensional metric
$$
ds^2=V^{-1} (d\tau+\omega)^2+ V ds^2_{(3)}\ ,
\eqn\lbone
$$
where
$$
ds^2_{(3)}={1\over q^2} \big(dx^2+dy^2+dq^2\big)\ ,
\eqn\lbtwo
$$
is the hyperbolic 3-metric
and 
$$
d\omega=\star dV
\eqn\lbthree
$$
with the Hodge duality operation taken with respect
 to the metric $ds^2_{(3)}$. The equation
\lbthree\ is just the magnetic monopole equation in
 a hyperbolic background.  The function $V$
is harmonic with respect to the hyperbolic 3-metric.  Solving
\lbthree\ for one monopole we get
$$
\eqalign{
V&=1+{1\over2} (\coth\rho-1)
\cr
\omega_x&=-{1\over2}{y\over x^2+y^2} \coth\rho
\cr
\omega_y&={1\over2}{x\over x^2+y^2} \coth\rho
\cr
\omega_z&=0\ ,}
\eqn\lbfour
$$
where
$$
\coth\rho={x^2+y^2+q^2+q_0^2\over \sqrt{(x^2+y^2+q^2+q_0^2)^2-4q^2 q_0^2}}\ .
\eqn\lbfive
$$
To construct the associated QKT geometry, let us suppose 
that the harmonic function $F$ depends
only on the coordinate $q$. Then we find that
$$
F=q^2\ .
\eqn\lbsix
$$
Therefore the associated QKT geometry is
$$
\eqalign{
ds^2_F&= q^2 \big[V^{-1} (d\tau+\omega)^2+ V ds^2_{(3)}\big]
\cr
H&=d\tau\wedge dx\wedge dy\ .}
\eqn\lbseven
$$
The metric $ds^2_F$ is the LeBrun metric which has been shown to be
complete in [\lebrun].  It is also known to 
have a non-vanishing but self-dual Weyl tensor.

\chapter{Twistor Spaces}

Let $M$  be a $4k$-dimensional weak QKT manifold.  Since 
the  connection $\Gamma$ of $M$ has
holonomy
$Sp(k)\cdot Sp(1)$, the tangent bundle is associated 
to a principal $Sp(k)\cdot Sp(1)$ bundle. In
particular this implies that the complexified tangent
 bundle $T_c M=T_{2k}\otimes T_2$ with the first 
subbundle associated with
$Sp(k)$ and the second associated with $Sp(1)$. 
\foot{In principle $T_2$ and $T_{2k}$ are  only
locally defined, but we shall assume that they exist
 globally for simplicity; this is similar to
demanding the existence of a spin structure and means
 that one can define a principal $Sp(k)\times
Sp(1)$ bundle. See [\sal] for a discussion.} Next we 
introduce a frame $e^{ai}$ and write the
metric as
$$
ds^2=\, e^{bj}\otimes e^{ai}\eta_{ab}\, \epsilon_{ij}\ , 
\eqn\trone
$$
where $\eta$ is the invariant $Sp(k)$ symplectic
 form ($a,b=1,\dots, 2k$) and $\epsilon$ is the
invariant $Sp(1)$ symplectic form ($i,j=1,2$). The
 reality condition for a vector $X$ in this
frame is
$$
\bar X_{ai}= X^{bj}\, \eta_{ba}\, \epsilon_{ji}\ ,
\eqn\trtwo
$$
which can be extended to tensors in a straightforward way.
A basis for the almost complex structures in this frame is
$$
(I_r)_{ai}{}^{bj}=-i \delta_a{}^b (\tau_r)_i{}^j\ ,
\eqn\trthree
$$
where the $\tau_r$ are the Pauli matrices; the
 almost complex structures are real tensors. 
The connection-form $\Gamma$ can be written in this basis as
$$
\Gamma_{ai}{}^{bj}=\delta_i{}^j A_a{}^b+\delta_a{}^b B_i{}^j\ ,
\eqn\connection
$$
where $A_a{}^b$ is the $Sp(k)$ connection and $B_i{}^j$ is
the $Sp(1)$ connection introduced in equation \inttwo. Similarly
 the curvature can be decomposed
as
$$
R_{ai}{}^{bj}=\delta_i{}^j F_a{}^b+\delta_a{}^b G_i{}^j\ .
\eqn\trfour
$$

The twistor space, $Z$, can be defined either as the
 projective bundle of $T_2$ or as the quotient
$U(1)\\ P$  of the principal $Sp(1)$ subbundle, $P$,
of the principal
$Sp(k)\times Sp(1)$ bundle. (We take the group of a
 principal bundle to act from the left.)
We shall work mainly  with $P$.  Functions on twistor
 space are $U(1)$ invariant
functions on $P$ while $U(1)$ equivariant functions
 on $P$ correspond to sections of  $U(1)$ line
bundles over $Z$ associated to $P$ considered as a
 $U(1)$ principal bundle over $Z$. This allows us
 to work with $P$ and then
reduce our results to $Z$.  For this we introduce
 ``coordinates'' $(x,u)$ on $P$, where $x$ are coordinates on the base
space $M$ and $u\in SU(2)$.  We write $u$ as $u_I{}^i$ 
(with inverse $u_i{}^I$), $i=1,2$ and $I=1,2$ with the local
$Sp(1)$ gauge transformations acting from the right, 
i.e. on the index $i$,  and the rigid
$Sp(1)$ transformations act from the left, i.e. on 
the index $I$, as we have already
mentioned. \foot{The equivariant formalism used here 
has been called  ``harmonic space'' formalism
elsewhere; it was applied to QK geometry in [\gio].}  
Since the structure group of $P$ as a
principal bundle over $Z$ is $U(1)$, it will be 
appropriate  to split up the capital $I$ indices
into two (1,2) indicating the $U(1)$ charges. The
right-invariant one-forms on the fibre (of
$P\rightarrow M$) in these coordinates are
$$
e_I{}^J=du_I{}^i u_i{}^J
\eqn\trfive
$$
with
$$
e_I{}^I=0\ 
\eqn\trsix
$$
as a consequence of the fact that $\det u=1$. 
The dual right-invariant vector fields $D_I{}^J$ satisfy
$$
D_I{}^Ju_K{}^i= \delta_K{}^J u_I{}^i-{1\over2} \delta_I{}^J u_K{}^i
\eqn\trseven
$$
and the algebra of vector fields is
$$
[D_I{}^J, D_K{}^L]=\delta_K{}^J D_I{}^L-\delta_I{}^L D_K{}^J
\eqn\treight
$$
which is isomorphic to the ${\cal L}Sp(1)$ Lie algebra.  To see this,
 we note that $D_I{}^I=0$ and set
$$
D_0=D_1{}^1-D_2{}^2\ .
\eqn\trnine
$$
It is then easy to verify that $\{D_0, D_1{}^2, D_2{}^1\}$
 satisfy the familiar Lie algebra
commutator relations of $SU(2)$.  We shall take the vector
 field $D_0$ to be tangent to  the orbits
of  $U(1)$ subgroup of $SU(2)$ acting on $P$ from the left
 which we have used to define the 
twistor space
$Z=U(1)\\ P$. We also note that
$$
\eqalign{
D_0 u_1{}^i&=u_1{}^i
\cr
D_0 u_2{}^i&=-u_2{}^i\ .}
\eqn\trten
$$
In the following we shall use the properties of the torsion
 and the curvature of the $Sp(1)$
connection
$$
\eqalign{
T_{a1b1c1}&\equiv u_1{}^i u_1{}^j u_1{}^k H_{aibjck}=0
\cr
G_{aibj,k\ell}&=\big(\epsilon_{ik}\epsilon_{j\ell}+
\epsilon_{i\ell}\epsilon_{jk}\big) G_{ab}\ ,}
\eqn\keyprop
$$
respectively. The latter condition holds provided
 that $k\geq 2$ and that $dH$ is (2,2)
 with respect
to all almost complex structures.  An outline of the proof of 
the above properties is given in the
appendix.

 Now we can state the properties of the twistor 
space $Z$ associated with a QKT manifold $M$:

\item(i)  $Z$ is a complex manifold provided that $k\geq 2$. 

\item(ii) $Z$ has a real structure.

\item (iii) $Z$ admits a complex contact structure
 provided that $k\geq 2$, $dH$ is (2,2) with
				respect to all almost complex structures and ${\rm det}(G_{ab})\not=0$.

\item (iv) $Z$ is a K\"ahler manifold provided that 
					    $(-\epsilon_{ij} G_{ab})$ is positive definite, 
$k\geq 2$ and $dH$ is (2,2) with
								respect to all almost complex
 structures as in the previous property.

The real structure is induced on $Z$ from the antipodal map
 on  each two-sphere fibre of $Z$ over
$M$ in exactly the same way as in hyper-K\"ahler  and
 quaternionic K\"ahler geometry, so  we refer
the reader to the literature for discussions of this point [\sal,\roc].

To prove (i) we introduce the horizontal lift basis on $P$:
$$
\eqalign{
\tilde E_I{}^J&=D_I{}^J
\cr
\tilde E_{aI}&= \hat e_{aI}-B_{aI,J}{}^K D_K{}^J}
\eqn\treleven
$$
with dual basis given by
$$
\eqalign{
E_I{}^J&=e_I{}^J+e^{aK} B_{aK, I}{}^J
\cr
E^{aI}&=e^{ai} u_i{}^I\ ,}
\eqn\trtwelve
$$
where we convert $i, j, k$ indices to $I, J, K$ indices using 
$u_I{}^i$ or $u_i{}^I$ as appropriate and where $\hat e_{ai}$ 
are the basis vector fields on $M$ dual to $e^{ai}$.  
We then find that
$$
\eqalign{
dE_I{}^J&=-E_I{}^K\wedge E_K{}^J + G_I{}^J
\cr
dE^{aI}&=-E^{aJ}\wedge E_J{}^I+ T^{aI}-E^{bI}\wedge A_b{}^a\ ,}
\eqn\torsion
$$
where $G_I{}^J$ is the $Sp(1)$ curvature  and $T^{aI}$ is the
torsion  in the $\{E_I{}^J, E^{aI}\}$ frame.  We claim that
 the set of vector fields
$\{\tilde E_{a1}, \tilde D_1{}^2\}$ spans an integrable
 distribution up to a $U(1)$ translation and
therefore defines a complex structure in
$Z$.  To show this, we write the second equation in
\torsion\ in the dual form
$$
[\tilde E_{aI},\tilde E_{bJ}]=- T_{aI bJ}{}^{cK}
 \tilde E_{cK}+ A_{aI, b}{}^c \tilde
E_{cJ}-A_{bJ,a}{}^c \tilde E_{cI}-G_{aIbJ,K}{}^L \tilde D_L{}^K\ .
\eqn\trthirteen
$$
Setting $I=J=1$ we find that the commutator $[\tilde E_{a1},
\tilde E_{b1}]$ closes on terms linear in
$\{\tilde E_{a1}, D_1{}^2\}$ and $D_0$ provided that
$$
\eqalign{
T_{a1b1}{}^{c2}&=0
\cr
G_{a1b1,1}{}^2&=0\ .}
\eqn\conditions
$$
Similarly,  the commutator $[\tilde E_{a1}, D_1{}^2]$ 
 closes on terms linear in $D_1{}^2$ and $D_0$.
The first condition in \conditions\ is equivalent to the
 first condition in \keyprop. For $k\geq 2$,
the second condition in \conditions\ is a special case of 
the second condition in \keyprop\ which holds for any weak QKT space, 
even if $dH$ is not (2,2)
with respect to all almost complex structures (see the appendix). 
Therefore for $k\geq 2$, the
twistor space
$Z$ is always a complex manifold. For $k=1$, the second
 condition in \conditions\ must be 
imposed in addition to
the conditions required by (4,0) local supersymmetry on 
the geometry of $M$. In particular, for the
examples that we have presented in section 3, this condition
 always holds because the Weyl tensor
is self-dual.

To show (iii), we first note that a complex contact structure
 is defined locally by a (1,0) form $\beta$ such
that
$$
\beta\wedge (\partial\beta)^k\not=0\ .
\eqn\contact
$$
In our case we choose
$$
\beta=E_1{}^2\ .
\eqn\trfourteen
$$
Using the definition of $E_1{}^2$ and the
 second condition in \keyprop, we find that
$$
dE_1{}^2=-E^{b2}\wedge E^{a2} G_{ab}-(E_1{}^1-E_2{}^2)\wedge E_1{}^2\ .
\eqn\twozero
$$
So
$$
\partial\beta=-E^{b2}\wedge E^{a2} G_{ab}\ ,
\eqn\trfifteen
$$
and the condition \contact\
is satisfied provided that
$$
det (G_{ab})\not=0\ .
\eqn\trsixteen
$$
As we have already mentioned, for $k\geq 2$ the
 second condition in \keyprop\ always holds
provided that $dH$ is (2,2) with respect to all
 almost complex structures. For $k=1$, the second
condition in \keyprop\ must be imposed in addition
 to those required by (4,0) local supersymmetry
on $M$.  For the examples that we have presented
 in section 3, this always holds since the
Weyl tensor is self-dual. Note that for $k=1$,
$dH$ is always (2,2) with respect to all almost complex structures.

It remains to show (iv). Since we have already shown
 that $Z$ is complex, it is enough to find the
appropriate K\"ahler form $\Omega$.  The metric can
 then be constructed from the K\"ahler form and
the complex structure. We choose as K\"ahler form 
$$
\Omega=2i \big(E_1{}^2\wedge E_2{}^1+ E^{b2}\wedge E^{a1} G_{ab}\big)\ .
\eqn\trkahler
$$
Clearly, $\Omega$ is (1,1) with respect to the chosen
 complex structure so it remains to show
that it is closed. For this, using \keyprop\ we find that
$$
\eqalign{
d\Omega&=2i\big(E^{c2}\wedge E^{b1}\wedge E^{a1} 
\nabla_{a1} G_{bc}+E^{c2}\wedge E^{b1}\wedge E^{a2} \nabla_{a2} G_{bc}
\cr &
+E^{c2}\wedge T^{b1} G_{bc}-T^{c2}\wedge E^{b1} G_{bc}\big)\ . }
\eqn\domega
$$
Expanding $T$  using the $E$ basis, we find that the
 term involving $E^{c2}\wedge E^{b1}\wedge E^{a1}$ in the above equation is
proportional to 
$$
\nabla_{a1}G_{bc}+{1\over2} T_{a1b1}{}^{d1} G_{dc}- T_{a1c2}{}^{d2} G_{db},
\eqn\trbbb
$$
antisymmetrised on $a$ and $b$, which vanishes
 because of the second Bianchi identity
$$
\nabla_{aI} G_{bJcK, LM} + T_{aIbJ}{}^{dN}
 G_{dNcK,LM}+{\rm cyclic\,\, in}\,\, (aI,bJ,cK)=0\ .
\eqn\trbbc
$$
Similarly, the term proportional to $E^{c2}\wedge E^{b1}\wedge E^{a2}$ 
in \domega\ vanishes and therefore
$\Omega$ is closed.    The metric is non-degenerate 
and positive definite provided that
$(-\epsilon_{ij} G_{ab})$ is non-degenerate and positive definite.

\chapter{HK structures from QKT manifolds}

As in the previous section, let $M$ be a $4k$-dimensional
 weak QKT manifold. As we have already
mentioned the tangent bundle can be written as 
$TM=T_{2k}\otimes T_2$.  The main task of this
section is to show that $\hat T_2$, which is
 defined to be $T_2$ with the zero section is
 removed, is a HK manifold provided that $k\geq 2$,
$dH$ is (2,2) with respect to all three almost
 complex structures and $(-\epsilon_{ij}G_{ab})$ is
non-degenerate and positive definite.  
Introducing complex coordinates $\{y^i; i=1,2\}$ along the fibres of
$\hT_2$, we define a set of $2k+2$ complex one-forms as follows:
$$
\eqalign{
E^i&=dy^i+y^j B_j{}^i
\cr
E^{a}&=e^{ai}y_i\ ,}
\eqn\hkone
$$
where $y_i=y^j\epsilon_{ji}$. We claim that this set
 of forms defines a complex structure on $\hT_2$, i.e., 
that it defines a basis set of (1,0) forms.
To show this we use the differential form version of
 Frobenius' theorem which states, in the
current context, that the exterior derivative of
 any (1,0) form should be a sum of two-forms each
one of which has a (1,0) factor. Differentiating \hkone\ we find 
$$
\eqalign{
dE^a&=-E^b\wedge A_b{}^a + e^{ai}\wedge  E_i+ T^{ai} y_i\cr 
dE^i&= -E^j\wedge B_j{}^i+ y^j G_j{}^i\ .}
\eqn\hktor
$$
Since
$H$ is (2,1) and (1,2) with respect to all almost complex structures, we can
write
$$
T_{aibjck}=
H_{aibjck} =\epsilon_{ij} H_{ab,ck}+\epsilon_{ki} H_{ca,bj}+\epsilon_{jk} H_{bc,ai}\ , 
\eqn\athree
$$
where $H_{ab,ck}=H_{ba,ck}$ and $H_{(ab,c)k}=0$, so that 
$$
T^{ai} y_i =2 e^{cj}\wedge E^b( H_{bc,}{}^a{}_j- H^a{}_{b, cj})\ .
\eqn\hkfour
$$
Then, using the expression in \keyprop\ for the $Sp(1)$ curvature $G$, we find
$$
y^j G_j{}^i=-e^{bi}\wedge E^a G_{ab}.
\eqn\hkfive
$$ 
Hence the right-hand sides of both of equations \hktor\ have
 the required structure for Frobenius' theorem to hold.

We
choose the first complex structure to be diagonal with
 respect to this integrable distribution,
i.e. $(IE)^i=iE^i$ and $(IE)^a=iE^a$. To find the
 metric and the rest of the hyper-K\"ahler
structure, it is enough to determine  two
of the three  K\"ahler forms, $\{\Omega_r,\ r=1,2,3\}$. As we are 
working in a basis in which  one of the
complex structures is diagonal, 
one of the K\"ahler forms, say $\Omega_1$, is a (1,1)-form 
with respect to the chosen complex
structure while the other two are (2,0) plus (0,2)
 with respect to the same complex
structure. 

The first K\"ahler form is
$$
\Omega_1=2i \big(\bar E_i\wedge E^i -\bar E_b\wedge E^a G_{a}{}^{b}\big)\ , 
\eqn\hkfour
$$
where the bars denote complex conjugation.\foot{Note
 that $Sp(1)$ and $Sp(k)$ indices 
are raised or lowered by complex conjugation 
as well as the corresponding symplectic invariant
tensors.} In particular, the frame $\{ \bar E_a, \bar E_i\}$ is
$$
\eqalign{
\bar E_i&=d\bar y_i-B_i{}^j\bar y_j \cr
\bar E_a&= -e_{ai}\bar y^i. }
\eqn\barframe
$$
The connection forms $\{B_i{}^j, A_a{}^b\}$ are
 skew-hermitian (e.g. ${\bar{(B_i{}^j)}}=\bar
B^i{}_j=-B_j{}^i$) and the basis forms $e^{ai}$
real with respect to the reality condition  \trtwo.
It is clear that $\Omega$ is (1,1)
with respect to the chosen complex structure, so 
it remains to show that it is closed.  That this
is so follows on using the second Bianchi
 identity for $G$, $DG_{ij}=0$, where $D$ is the $Sp(1)$
covariant exterior derivative, and contracting it with $y^i \bar y^j$.

Next we choose the second almost complex structure $J$ to be
$$
\eqalign{
J(E^a)&=\bar E_b\eta^{ba}\ , \qquad J(\bar E_a)=E^b \eta_{ba}
\cr
J(E^i)&=\bar E_j\epsilon^{ji}\ ,\qquad J(\bar E^i)=E^j\epsilon{ji}\ .}
\eqn\secondcompl
$$
The almost complex structure $J$ is integrable
as may easily be seen by observing that a 
 basis of (1,0) forms for $J$ is
$\{E^a+i\bar E^a, E^i+i\bar E^i\}$ and then
 by using  the Frobenius' theorem. The $J$ complex
structure anticommutes with the $I$ complex 
structure as required.   The (2,0) part of the
K\"ahler form of the $J$ complex structure is 
$$
\Omega'= E^j\wedge E^i\epsilon_{ij}- E^b\wedge E^a G_{ab}\ .  
\eqn\hkfive
$$
The proof that this form is
closed is similar to that for $\Omega_1$ with the
 difference that one must use the second Bianchi
identity for $G$ contracted with $y^i y^j$.
 The K\"ahler forms $\Omega_2, \Omega_3$ are the
real and imaginary parts of $\Omega'$, 
$$
\Omega'={1\over2}(\Omega_2 + i\Omega_3)\ .
$$
This shows that $\hat T_2$ is an HK
manifold since the third complex structure
 can be constructed form the first two and its
integrability is also implied by the 
integrability of the first two. The metric is
$$
ds^2= 2\bar E^i\otimes E_i -2\bar E^b\otimes E^a G_{ab}. 
\eqn\hksix
$$
It is hermitian with respect to all three 
complex structures and is non-degenerate and
positive definite provided that
$-\epsilon_{ij}G_{ab}$ is non-degenerate and positive definite.

\chapter{Concluding Remarks}

We have shown that the QKT geometry of 
 manifolds  that arise  in the context
of two-dimensional (4,0) locally supersymmetric
 sigma models  is determined by the
properties of a metric connection with torsion. 
This connection has holonomy $Sp(k)\cdot
Sp(1)$ so that the corresponding geometry
 is a generalization of QK geometry.  
QKT manifolds admit a twistor construction.  
The twistor space  is a $(4k+2)$-dimensional
K\"ahler manifold with a complex contact structure.  
In addition, for every QKT manifold there is a
$(4k+4)$-dimensional hyper-K\"ahler manifold which
 is obtained from a vector bundle over the QKT manifold with fibre
$\bC^2$ associated to the $Sp(1)$ principal
 bundle by omitting the zero section. 

There are various limits in the twistor 
construction for QKT manifolds in which one or more tensors
associated with this structure vanish. In the 
limit that the torsion vanishes, as we have already
mentioned, the QKT structure degenerates
 to a QK one and one recovers the results of
Salamon [\sal] and Swann [\swann] for QK manifolds. 
In another limit where the torsion
does not vanish but the holonomy becomes $Sp(k)$ 
the manifold becomes HKT for which the  
twistor construction was given in [\hkt].  Finally, 
if both the torsion vanishes and the holonomy
is $Sp(k)$, then the manifold is HK for which the 
twistor construction was given in [\roc].

In this paper we have not investigated the 
applications of the twistor construction  in (4,0) supergravity
coupled to sigma model matter system. However, 
it is likely that the sigma model maps can be
thought  as holomorphic maps from a harmonic
extension of the (4,0) superspace to the
 twistor space of $M$,  thus generalizing a  similar
property of (4,0) superfields for the models 
with rigid supersymmetry [\dks, \hkt].

It would also be of interest to find more
 examples of QKT manifolds in $4k$-dimensions for $k>1$.
For example, there might be locally symmetric 
spaces with a QKT structure in direct analogy to the
Wolf spaces for QK manifolds [\wolf] or to the
 group manifold examples for HKT manifolds [\vp]. New
QKT manifolds may also be constructed starting from  
QK manifolds with an isometry that respects the
QK structure and then performing a T-duality 
transformation along the Killing direction. By this means one might expect to 
develop relationships between QK and QKT manifolds similar to those that 
hold between HK and HKT manifolds [\gps, \ao].

\vskip 1cm
\noindent{\bf Acknowledgments:} One of us G.P. would 
like to thank R. Goto for helpful
discussions.  A.O. is supported by EPSRC and the German
 National Foundation. G.P. is supported by a
University Research Fellowship from the Royal Society.
\vskip 1cm

\appendix

Here we shall show that
$$
\eqalign{
T_{a1b1}{}^{c2}&=0
\cr
G_{aibj,k\ell}&=(\epsilon_{ik} \epsilon_{j\ell}
+\epsilon_{i\ell}\epsilon_{jk}) G_{ab}\ . }
\eqn\algebraic
$$
The first condition is equivalent to 
$$
T_{a1b1c1}\equiv u_1{}^i u_1{}^j u_1{}^k H_{aibjck}=0\ .
\eqn\atwo
$$
 Then the first condition in \algebraic\ follows
by contracting the expression for $H$ 
in \athree\ with $u_1{}^i$ as in \atwo.

To show the second condition in \algebraic,
 one uses the Bianchi identity
$$
R_{\mu [\nu\rho\sigma]}={1\over3} \nabla_{\mu}
 H_{\nu\rho\sigma}-2 P_{\mu\nu\rho\sigma}\ ,
\eqn\bianchi
$$
where $P_{\mu\nu\rho\sigma}=3\partial_{[\mu} H_{\nu\rho\sigma]}$.
We first write this Bianchi using the $ai$ 
coordinate description and then contract all four
 $Sp(1)$ indices $i,j,k,\ell$ with 
$u_1{}^i, u_1{}^j, u_1{}^k, u_1{}^\ell$.  This gives
$$
R_{a1[b1c1d1]}={1\over3} \nabla_{a1} T_{b1c1d1}-2 P_{a1b1c1d1}\ . 
\eqn\bianchib 
$$
From the first condition in \algebraic, we 
find that the left-hand-side of \bianchib\
vanishes.  But
$$
R_{aibj,ckd\ell}=\epsilon_{k\ell} F_{aibj,cd}+\eta_{cd} G_{aibj,k\ell}
\eqn\curvdec
$$
and so we find that
$$
R_{a1b1c1d1}=\eta_{cd} G_{a1b1,11}\equiv \eta_{cd} G'_{ab}\ ,
\eqn\afour
$$
where $F_a{}^b$ is the curvature of the $Sp(k)$ connection $A_a{}^b$ in \connection. 
Substituting this in \bianchib\ we find that
$$
G'_{ab}=0\ ,
\eqn\afive
$$
provided that $k\geq 2$. Since $G'_{ab}=0$ for any $u$
 and $G'_{ab}=u_1{}^i\, u_1{}^j\, u_1{}^k\,
 u_1{}^\ell\, G_{aibj,k\ell}$, this implies that
$$
G_{ab, (ijk\ell)}=0\ .
\eqn\tsymm
$$
We remark that this condition is enough to
 show that the twistor space is a complex manifold.

Next we contract the Bianchi identity 
 \bianchi\ with $u_2{}^i, u_1{}^j, u_1{}^k, u_1{}^\ell$ and
we get
$$
R_{a2[b1,c1d1]}={1\over3} \nabla_{a2} T_{b1c1d1}-2 P_{a2b1c1d1}\ . 
\eqn\bianchic
$$
The right-hand-side of the above equation vanishes 
provided that $P_{a2b1c1d1}=0$ which is
precisely the condition for $dH$ to be (2,2) with
 respect to all three almost complex structures.
Using \curvdec\ for the curvature, we find that
$$
R_{a2b1,c1d1}= \eta_{cd}\, G_{a2b1,11}\ .
\eqn\asix
$$
Substituting this back into \bianchic, we find that
 both $G_{(ab)21,11}$ and $G_{[ab]21,11}$ vanish
provided that $k\geq 2$. Since this is again the case
 for any $u$, the first condition implies
that $G_{(ab)ij}$ vanishes and the second together
 with \tsymm\ imply the second condition in
\algebraic. We remark that for QK manifolds
 $G_{ab}=\lambda\, \eta_{ab}$ where $\lambda$ is a real
constant.

\refout

\bye